\documentclass[12pt]{iopart}
\usepackage{epsfig}
\usepackage{iopams, setstack}
\usepackage{epsfig,psfrag}
\usepackage{
amssymb}
\usepackage{exscale}
\usepackage{mathrsfs}
\usepackage{times}
\usepackage[T1]{fontenc}

\newcommand{\ket}[1]{\left| #1 \right\rangle}

\begin{document}
\title[The Aharonov-Bohm effect: the role of tunneling and associated
forces]{The Aharonov-Bohm effect: the role of tunneling and associated
forces} 

\author{G.~C.~Hegerfeldt and J.~T.~Neumann}
\address{Institut f\"ur Theoretische Physik, Universit\"at
G\"ottingen, Friedrich-Hund-Platz 1, 37077 G\"ottingen, Germany}
 
\date{Received: date / Revised version: date}
\eads{\mailto{hegerf@theorie.physik.uni-goettingen.de},
  \mailto{j.t.neumann@gmx.de}}


\begin{abstract}
Through tunneling, or barrier penetration, small wavefunction tails
can enter  a finitely 
shielded cylinder with a  magnetic field inside. When the shielding increases
to infinity the Lorentz force goes to zero together with these
tails. However, it is shown, by considering the radial derivative of
the  wavefunction on the cylinder surface, that a flux dependent force
remains.  This force explains in a natural way the Aharonov-Bohm
effect in the idealized case of infinite shielding.
\end{abstract}



\submitto{\JPA}

\section{Introduction}
The counterintuitive Aharonov-Bohm (AB) effect \cite{ab1959} represents one of
the most widely discussed issues of quantum physics.  It predicts
that a charged particle can be influenced by a magnetic field ``even
if the particle is nowhere in the region of nonzero field strength''
\cite{an2002}.  This intriguing effect has alternatively been
attributed to a nonlocal feature of quantum mechanics, to a hitherto
unexpected direct physical meaning of an otherwise unphysical vector
potential, to a topological cause and so on \cite{pt1989}.  It is
experimentally well verified \cite{pt1989,to1986a},  and its possible
applications have attracted much interest recently,  see \emph{e.g.}
\cite{ha2004,ls2000,ba1999,emacd2004,za2004,co2004}. A generally
accepted intuitive and 
physical understanding, however, seems to be still lacking.  

\begin{figure}[h]
\begin{center}
\epsfig{file=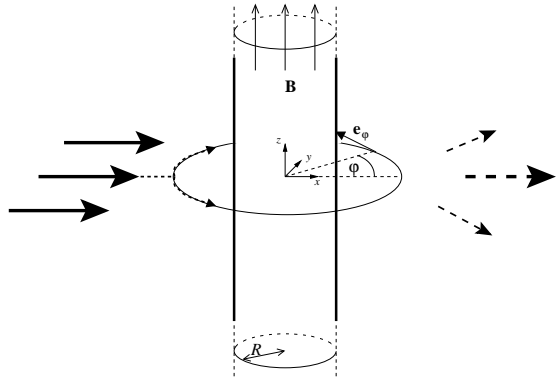, width=0.47\textwidth}
\caption{\textbf{Schematic sketch of the Aharonov-Bohm effect.}
  Electron scattering by an infinite, perfectly shielded cylinder
  of radius $R$, with a homogeneous magnetic field $B \mathbf{e}_z$
  inside, depends on the magnetic flux.}
\label{setup_fig}
\end{center}
\end{figure}

As Hamiltonian operator for an electron (without spin) in the presence
of  a magnetic field one  takes
\begin{equation}\label{H}
\hat H = \left [\hat{\mathbf P} + e  \mathbf A(\hat{\mathbf
    x})\right]^2/2m_e 
\end{equation}
where $\hat{\mathbf P}$ is the canonical momentum operator of the
electron, $m_e$ the electron mass and $-e$ its charge.  $\mathbf
A(\mathbf x)$ is a vector potential, which is not unique and is
 related to the magnetic field by $\mathbf
B(\mathbf x) = \mathrm{curl} \, \mathbf A(\mathbf x)$.  In classical
physics such a vector potential is an auxiliary quantity without
direct physical meaning. It may happen that a magnetic field vanishes
in a region while the allowed vector potentials do not. An example is
given in fig.~\ref{setup_fig}
with a constant magnetic field which
vanishes outside an infinitely long cylinder of radius $R$.
Since  the
integral  $\oint\mathbf A(\mathbf x)\cdot {\rm d} \mathbf x$ over a circle
around the cylinder yields the flux, $\mathbf
A(\mathbf x)$  cannot identically
vanish outside the cylinder.


A quick, heuristic, way to obtain the AB effect is to note that an
electron may pass on either side of an impenetrable cylinder (see
fig.~\ref{setup_fig}), thereby picking up a phase $ (e/\hbar) \int
\mathbf A(\mathbf x) \cdot {\rm d} \mathbf{x}$, with the integral taken
over the electron's path.  The resulting phase difference is $(
e/\hbar)\Phi$, where $\Phi$ is the magnetic flux in the cylinder, and
this leads to field dependent  effects for
scattering. Related  effects in magnetic fields without
shielding had been pointed out earlier \cite{fr1939,es1949}.

A more detailed derivation \cite{ab1959,kr1965b} describes
electron scattering  by eigenfunctions of the Hamiltonian
(\ref{H}) with $\mathbf A(\mathbf x)= (\Phi/2\pi r)\mathbf
{e}_\varphi$, where
$r\equiv\sqrt{x^2+y^2}\ge R$.
Impenetrability is taken into account by choosing eigenfunctions with
zero boundary conditions. Since the vector potential outside does not
vanish, the eigenfunctions differ from those for the free Hamiltonian
$\hat{H}^{(0)}=\hat {\mathbf{P}}^2/2m_{\rm{e}}$ with the same boundary
conditions. This then gives the baffling result that electron
scattering depends on the magnetic field inside the cylinder although
the electron wavefunction cannot penetrate there.  Ref. \cite{pt1989}
explicitly calculates for a special case the 
momentum transfer to the electron from the wall of the excluded
cylinder, compares it with the momentum transfer implied by the
scattering cross section, and concludes that in general the force
exerted by the cylinder surface is needed to satisfy Ehrenfest's
theorem. It remains, however, physically somewhat puzzling why this
force depends on the magnetic field inside the cylinder although the
wavefunction vanishes on the surface and
inside. Ref. \cite{Liebowitz65} invokes a new, previously not
considered, classical force to explain the AB effect. This was
criticized in \cite{Hrasko66} and rebutted in
\cite{Liebowitz66}. Ref. \cite{Boyer2000} draws a distinction between
the Aharonov-Bohm phase shift and the Aharonov-Bohm effect and
suggests that the Aharonov-Bohm phase shift is actually due to
classical electromagnetic forces when relativistic effects are taken
into account. In a recent experiment \cite{Batelaan2007} the time
delay for an electron passing by a ``macroscopic'' solenoid has been
investigated. 

Some authors have noted that if one starts right away with a perfectly
shielded cylinder -- corresponding to quantum mechanics in a plane
with a hole or in three-dimensional space with an excluded cylinder --
the quantum 
dynamics for a particle outside is not uniquely determined ({\it cf.},
{\it e.g.}, \cite{mv1995}). In this point of view the choice of one
particular quantum dynamics then appears somewhat {\it ad hoc} since
it typically relies on 
information from the excluded region.  This problem does not arise in
approaches which consider finite, but increasingly high, barriers and
then take a limit \cite{kr1965b,mv1995} .
In this case there are no interpretational problems. But it has also
been shown 
\cite{kr1965b,mv1995} that as the barrier height is increased
in the limit one arrives at exactly the same zero 
boundary conditions for the wavefunction as before. Thus the
conceptual problem remains by which physical mechanism 
information about the inside field is transmitted to the outside. 

In this paper it is shown that in case of large, but finite, shielding
the small wavefunction tails, which can tunnel into the cylinder and
into the magnetic field, determine the force acting on the
electron. In the limit of infinite shielding this force remains finite
and flux dependent, it determines the otherwise under-defined dynamics
outside and yields the AB effect. 

The paper is organized as
follows. In section~\ref{sec_indeterminacy} we review the argument why
the quantum dynamics can be viewed as ambiguous if one  has a perfectly
shielded cylinder right from the beginning. This is due to the fact
that classically equivalent  
Hamiltonians can become physically inequivalent after transition to
quantum mechanics. In section~\ref{sec_forces}
expressions for the forces for finite and infinite shielding are given
in terms of radial derivatives  of the modulus of the
wavefunction.  These radial derivatives are calculated in
section~\ref{sec_radial} and the resulting forces are determined. In
section~\ref{sec_forces_and_AB} it is indicated how the forces
determine the dynamics and the AB effect. In the appendix expressions
for the forces are explicitly derived which are valid for arbitrary
geometries, not only cylinders.


\section{Indeterminacy of the quantization}
\label{sec_indeterminacy}
 We first
consider a free {\em classical} electron 
outside an impenetrable  infinitely long cylinder of
radius $R$. No assumption about a
possible magnetic field inside the cylinder is made.
By symmetry one can confine oneself to the $x-y$ plane. A possible
classical Hamiltonian function is $H^{(0)} = 
{\mathbf{P}}^2/2m_{\rm{e}}$, with reflections from the boundary of the
excluded region
implemented by the configuration space. The Hamiltonian, however, is
not unique. Indeed, if $\mathbf \Omega_{\rm dum}(\mathbf x)$ is any
vector function (a ``dummy field'') outside the cylinder, satisfying 
$$
{\rm curl}\,\mathbf \Omega_{\rm dum}(\mathbf x) =0
$$ 
there, otherwise
arbitrary and not connected to any field inside the cylinder,
then 
\begin{equation}\label{dummy}
 H_{\rm dum} =\left[ {\mathbf P} +   \mathbf \Omega_{\rm   dum}({\mathbf
     x}) \right]^2/2m_{\rm{e}} 
\end{equation}
equals the kinetic energy and also yields\footnote{This corresponds to
the well known freedom to add to the Lagrangian a total time
derivative. {\it Cf.}, \emph{e.g.}, \cite{mv1995}.}  $\ddot{\mathbf{x}}=0
$.
This is, up to a constant, the most general form of the Hamiltonian
yielding $\ddot{\mathbf{x}} = 0$ and equaling the kinetic energy.
Classically, all these Hamiltonians are physically equivalent. 
\begin{figure*}[htb]
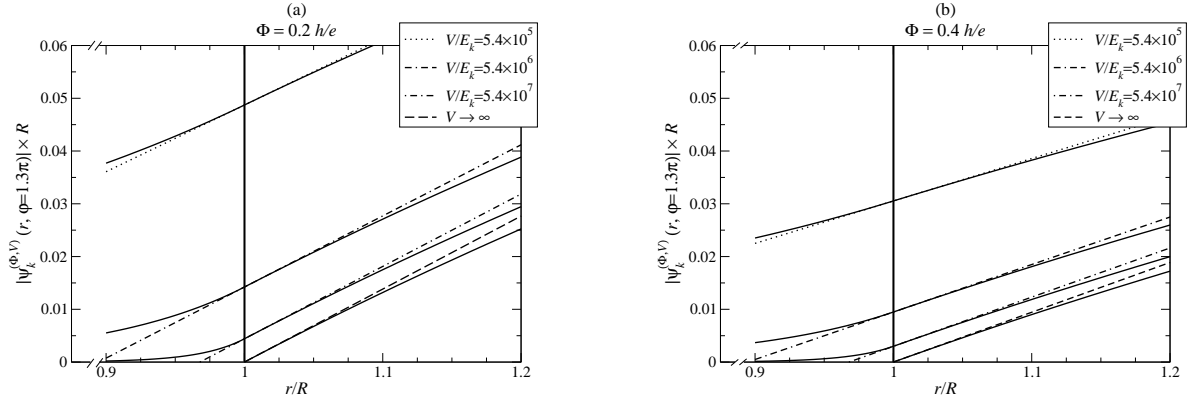

\begin{center}
\begin{minipage}{0.45\textwidth}
\begin{center}
\epsfig{file=hegerfeldt_fig2a.eps, width=\textwidth}
\end{center}
\end{minipage}
\hfill
\begin{minipage}{0.45\textwidth}
\begin{center}
\epsfig{file=hegerfeldt_fig2b.eps, width=\textwidth}
\end{center}
\end{minipage}
\caption{\textbf{Wavefunction behavior near the boundary: finite
  shielding.}  Cylinder of radius $R$,  wavefunction 
  corresponding to incoming  
  electrons of momentum $\hbar k \mathbf{e}_x$ with $kR = 4.3 \times
  10^{-3}$. Two different 
  magnetic fluxes: (a) $\Phi =   0.2 \, h/e$, and (b) $\Phi = 0.4 \, h/e$. 
 Solid lines: $|\psi_k^{(\Phi,V)}(r, \varphi)|$ near the 
  cylinder boundary  for fixed polar angle $\varphi = 1.3 \pi$  and for
  increasing barrier heights $V$ 
  as well as for the limit $V \rightarrow \infty$. Thick vertical line:
  cylinder boundary. Broken lines: 
  slope at this point. For the two fluxes the slopes differ
  significantly, but differ little for a fixed flux and different  barrier
  heights $V$.}  
\label{slope_fig}
\end{center}
\end{figure*}
\begin{figure*}[htb]
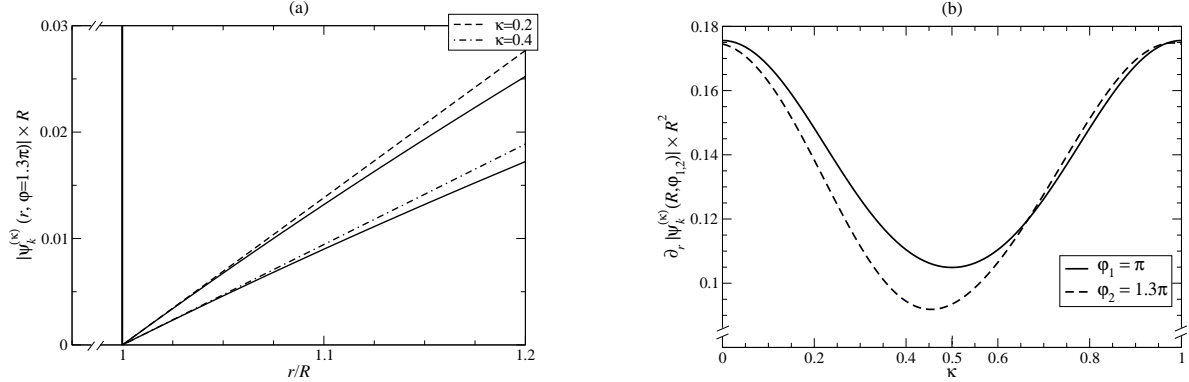

\begin{center}
\begin{minipage}{0.45\textwidth}
\begin{center}
\epsfig{file=hegerfeldt_fig3a.eps, width=\textwidth}
\end{center}
\end{minipage}
\hfill
\begin{minipage}{0.45\textwidth}
\begin{center}
\epsfig{file=hegerfeldt_fig3b.eps, width=\columnwidth}
\end{center}
\end{minipage}
\caption{\textbf{Infinite
  shielding with dummy field. 
}
$kR$ as in fig. \ref{slope_fig}. (a) Solid lines:
  $|\psi_k^{(\kappa)}(r,\varphi)|$  near the   boundary  as a function
  of $r$ for fixed polar 
  angle $\varphi=1.3 \pi$ and for two different values of  $\kappa$
  which labels inequivalent dummy fields 
  $\mathbf{A}_\mathrm{dum}^{(\kappa)}$ from  ~(\ref{kappa}). The
  slope at $r=R$ (dashed, dot-dashed lines)  depends on   $\kappa$.  
(b) The slope at $r=R$  plotted as   a function of  $\kappa$  for two
polar angles  
  $\varphi_1 = \pi$ and $\varphi_2 = 1.3 \pi$.  The value of $\kappa$
  is uniquely determined if the slope is known for two different
  angles $\varphi$.}
\label{kappa_fig}
\end{center}
\end{figure*}
This general equivalence, however, is no longer true after transition
to the quantum theory  (with zero boundary conditions
on the cylinder) ({\it cf.}, {\it e.g.}, \cite{mv1995,ru1983}). 
For a given $\mathbf{\Omega}_\mathrm{dum} (\mathbf{x})$ with
$\mathrm{curl}~\mathbf{\Omega}_\mathrm{dum} (\mathbf{x}) = 0$ outside the
cylinder, we define $\kappa$ as
\begin{equation}
\kappa \equiv {\rm non-integer~part~ of~ } \oint \mathbf{\Omega}_\mathrm{dum}
(\mathbf{x}) \cdot {\rm d} \mathbf{x} / h
\label{kappa_def}
\end{equation}
so that $0 \leq \kappa < 1$.  Then one can define the
continuous and differentiable function of
 modulus 1 
\begin{equation}
\Lambda (\mathbf{x}) \equiv \exp \left\{ \frac{i}{\hbar}
  \int_{\mathbf{x}_0}^\mathbf{x} \left[ \mathbf{\Omega}_\mathrm{dum}
  (\mathbf{x}') - \frac{\hbar \kappa}{r'}  \mathbf{e}_{\varphi'}
  \right] \cdot {\rm d} \mathbf{x}' \right\}
\label{transform}
\end{equation}
where $\mathbf{x}_0$ can be chosen as $\mathbf{x}_0 = (R,0)$;
$r,~r'\ge R$.  It is easy to check that $\Lambda(\hat{\mathbf{x}})
\hat{H}_{\rm dum} \Lambda (\hat{\mathbf{x}})^\dagger = \hat{H}_{\rm
dum}^{(\kappa)}$ where
\begin{equation}\label{kappa}
 \hat{H}_{\rm   dum}^{(\kappa)} =\left[
 \hat{{\mathbf P}} +   \mathbf \Omega^{(\kappa)}_{\rm   dum}(\hat{{\mathbf x}})
\right]^2/2m_{\rm{e}}, ~~~~
  \mathbf \Omega^{(\kappa)}_{\rm dum}(\mathbf{x})\equiv  \frac 
  {\hbar\kappa}{r}\,\mathbf   {e}_\varphi~
\end{equation}
with $0 \leq \kappa < 1$,
and the zero boundary conditions are preserved. The
 Hamiltonian operators $\hat{H}_{\rm   dum}^{(\kappa)}$
are physically inequivalent for  $0 \leq \kappa < 1$, and without
further input information it is unclear which one to choose. Thus
there is an ambiguity when one starts right away with infinite
shielding, and this ambiguity can be fixed by imposing an 
additional boundary condition which arises in a natural way when one
takes into account tunneling and associated forces, as seen in the following.

\section{The forces for finite and infinite shielding}
\label{sec_forces}

Finite shielding of the cylinder can be modeled by a 
potential $V (\mathbf{x})$. Just for simplicity we consider $V
(\mathbf{x}) = V_0 \Theta (R-r)$, with  $V_0 \rightarrow \infty$
later on, and a homogeneous magnetic field  inside the cylinder with
vector potential 
\begin{equation}
\mathbf{A}_\Phi (\mathbf{x}) = \frac{\Phi}{2 \pi} \left[ \frac{1}{R^2}
  \Theta (R-r) + \frac{1}{r^2} \Theta (r-R) \right] \left(
  \begin{array}{c} -y \\ x \end{array} \right)
\end{equation}
where $\Phi$ is the magnetic flux, but also other magnetic fields can be
considered. The Hamiltonian then is 
\begin{equation}
\hat{H}_{\Phi,V} = \left[ \hat{\mathbf{P}}^2 + e \mathbf{A}_\Phi
  \left(\hat{\mathbf{x}}\right) \right]^2/2m_{\rm{e}} + V
\left( \hat{\mathbf{x}} \right).
\end{equation}
We consider a normalized (planar) wavefunction, $\psi_t^{(\Phi,V)}$, 
corresponding 
for $t \to -\infty$ to an incoming free particle which is then
scattered. Due to tunneling small tails enter the cylinder. The total
force on the electron is
\begin{equation}\label{force0}
\mathbf{F}^{(\Phi,V)}_t = \langle
\psi_t^{(\Phi,V)}|-\nabla V - e\hat{\mathbf{v}} \times
\hat{\mathbf{B}}|\psi_t^{(\Phi,V)}\rangle ~.
\end{equation}
Note that only the tails contribute. As the tails, the Lorentz force
goes to 0 when $V \to \infty$. For a step potential the first part
becomes $V_0 \int_0^{2\pi}R {\rm d} \varphi \,
|\psi_t^{(\Phi,V)}(R,\varphi)|^2\mathbf{e}_r$. Expanding
$\psi_t^{(\Phi,V)}$ in terms of eigenfunctions of $\hat{H}_{\Phi,V}$
and using eqs.~(35) - (40) of \cite{kr1965b} one can show explicitly
for large $V_0$ that $V_0^{1/2} |\psi_t^{(\Phi,V)}(R,\varphi)|= \hbar
\partial_r |\psi_t^{(\Phi,V)}(R,\varphi)|/(2m)^{1/2} +
O(V_0^{-1})$. Thus
\begin{equation}\label{force1}
\mathbf{F}^{(\Phi,V)}_t = \frac{\hbar^2}{2m_{\rm{e}}}\int_0^{2\pi}R
{\rm d} \varphi \left[\frac{\partial}{\partial 
    r}|\psi_t^{(\Phi,V)}(R,\varphi)|\right]^2\mathbf{e}_r+ O(V_0^{-1}). 
\end{equation}
This result is also true for more general $V$ and $\mathbf{B}$,
\emph{e.g.} $\mathbf{B}(r)=0,~r>R/2$, as well as for the torus and other
domains (with a surface integral and normal derivative). This general
case is treated in the appendix.

For infinite shielding and fixed arbitrarily chosen dummy field $\mathbf
\Omega^{(\kappa)}_{\rm dum}$ we use $\mathbf{F}^{(\kappa)}_t =
{\rm d}/{\rm d} t \,
\langle\psi_t^{(\kappa)}|m_{\rm{e}}\hat{\mathbf{v}}|\psi_t^{(\kappa)}\rangle$
for the force and eventually obtain
\begin{equation}\label{force2}
\mathbf{F}^{(\kappa)}_t = \frac{\hbar^2}{2m_{\rm{e}}}\int_0^{2\pi}R
{\rm d} \varphi \, \left[\frac{\partial}{\partial
    r}|\psi_t^{(\kappa)}(R,\varphi)|\right]^2\mathbf{e}_r. 
\end{equation}
Again this carries over to other domains as shown in the appendix.

The as yet underdetermined dummy field
$\mathbf{\Omega}_\mathrm{dum}^{(\kappa)}$ can be made unique by
requiring that the $\kappa$-dependent force (\ref{force2}) agrees with
(\ref{force1}) in the limit $V \to \infty$ (obtained with the field
actually contained in the cylinder). It will now be shown that this
requirement uniquely determines $\kappa$, and thus also the dynamics
of the idealized case.

\section{The radial derivative and resulting forces}
\label{sec_radial}

It suffices to consider electron beams of
definite incoming kinetic momentum, $\hbar k \mathbf e_x$ say. The
corresponding scattering solution is  a stationary state, denoted by
$\psi_k^{(\Phi,V)}$ and $\psi_k^{(\kappa)}$, respectively. To
calculate these we note that around the backward
direction, $\varphi = \pi$, the incident wave
should  behave as an eigenfunction of $\hat{\mathbf
  P}_\mathrm{kin}$.  This implies that the incident wave should behave as
\begin{equation}
e^{i \mathbf{k} \cdot
    \mathbf{x} - i e\Phi ( \varphi - \pi)/h}~~\mathrm{for}~
  kr
 \gg 1,~| \varphi  - \pi | <
  \varepsilon~
\label{asymp}
\end{equation}
and similarly for $\psi_k^{(\kappa)}$. The asymptotics in
Ref. \cite{ab1959}, based on the probability current, is the
same\footnote{One can also start with an incident plane wave
$\ket{\mathbf{k}} \hat{=} e^{i \mathbf{k} \cdot \mathbf{x}}$ and use
the Lippmann-Schwinger equation to obtain the corresponding scattering
solution. We have determined the necessary Green's functions and have
shown that one obtains the same expression as with
 ~(\ref{asymp}). This generalizes
a result for a magnetic string \cite{so1997} and  will be published
elsewhere.}.

The wavefunction $\psi_k^{(\Phi,V)}(r,\varphi)$ can be calculated
analytically  as a series and evaluated numerically. 
Note  that the wavefunction is gauge dependent while its absolute
 value is not. Although the tails   
  go to zero inside and on  the cylinder   when  $V$ tends to infinity,
 the behavior of the wavefunction in the vicinity of the 
cylinder  depends sensitively on the value of
$\Phi$, as seen in figs.~\ref{slope_fig} (a) and \ref{slope_fig}
(b). There
we consider two different magnetic fields inside the cylinder, with
fluxes $\Phi_1$ and $\Phi_2$, and increasing barrier heights $V$. For
each flux and barrier height 
$\psi_k^{(\Phi,V)}(r,\varphi)$ is calculated  and its
absolute value is plotted as a function of $r$ for $\varphi = 1.3\,
\pi$. With increasing barrier height $V$ the wavefunctions are indeed
seen to converge to zero inside the cylinder ($r \leq R$). 
The  form of the tails depends on the specific $\mathbf{B}$ and V, while
the limiting slope depends only on $\Phi$, or rather on 
\begin{equation}\label{alpha}
 \alpha\equiv {\rm non-integer~ part ~ of~ } e\Phi/h.
\end{equation}

With infinite shielding and a dummy field $\mathbf \Omega^{(\kappa)}_{\rm
  dum}$, the scattering solutions
$\psi_k^{(\kappa)}(r,\varphi)$ vanish 
  on the cylinder ($r=R$), but  the rate with which 0
  is approached when $r \to R$ depends on $\kappa$. To see this we
  calculate $\psi_k^{(\kappa)}(r,\varphi)$ by expressing it as a linear
combination of eigenfunctions of ${\hat H}^{(\kappa)}_\mathrm{dum}$ of
fixed canonical angular momentum, with unknown
coefficients. Using the asymptotics of Bessel and Hankel functions and
  ~(\ref{asymp}) one can determine the 
coefficients and obtains
\begin{eqnarray}
\psi_k^{(\kappa)} (r,\varphi) & = & \sum_{n=-\infty}^\infty
(-i)^{|n+\kappa|} e^{in (\varphi + \pi)} \nonumber \\ 
& & \times \left[
J_{|n+\kappa|} (kr) - \frac{J_{|n+\kappa|} (kR)}{H_{|n+\kappa|} (kR)}
H_{|n+\kappa|} (kr) \right]~.
\label{scattering}
\end{eqnarray}
The rate is given by the tangent slope of
  $|\psi_k^{(\kappa)}(r,\varphi)|$  at the cylinder, \emph{i.e.} by the
  radial derivative $\partial 
  _r |\psi_k^{(\kappa)}(r=R,\varphi)|$ which can  be  calculated from
   ~(\ref{scattering}). Since $\psi_{k}^{(\kappa)} = 0$ on
  the boundary of the cylinder one easily sees that
\begin{equation}
\partial_r \left| \psi_k^{(\kappa)}(r,\varphi) \right|=\left|
  \partial_r \psi_k^{(\kappa)} (r,\varphi) \right|   \quad
\mathrm{if} \quad r=R.
\end{equation}
With this identity and using the Wronskian for Bessel and Hankel
functions, one obtains from  ~(\ref{scattering})
\begin{equation}\label{derivative}
\partial_r \left| \psi_k^{(\kappa)}(R,\varphi) \right| =
\frac{2}{\pi R} \left| \sum_n
\frac{(-i)^{|n+\kappa|}}{H_{|n+\kappa|}(kR)} e^{in (\varphi + \pi)} \right|.
\end{equation}
We note that this derivative is invariant under the unitary
transformation $\Lambda(\hat{\mathbf{x}})$ in  ~(\ref{transform}).

In fig.~\ref{kappa_fig} (a) $|\psi_k^{(\kappa)}(r,\varphi)|$ is
  plotted as a function of $r$ for two values of $\kappa$ and for
  fixed $\varphi=1.3\pi$ and it is seen that the slope at $r = R$
  depends on $\kappa$.  
In fig.~\ref{kappa_fig} (b) the slope at $r=R$ is
plotted as a function of $\kappa$  for two
values of $\varphi$ (solid curve: $\varphi_1 =\pi$, dashed curve:
$\varphi_2 = 1.3 \pi$) and it is seen that the values of the slope
at the two different angles determine $\kappa$ uniquely. Moreover,
this slope agrees with the limiting slope for a magnetic flux $\Phi$
if  $\kappa$ equals the non-integer part of $e\Phi/h$
(\emph{i.e.} if $\kappa=\alpha$), and then the corresponding forces on the
electron beam are equal, by eqs.~(\ref{force1}) and (\ref{force2});
conversely  the condition $\kappa=\alpha$ is also necessary for this.
Thus requiring the equality of the forces determines the previously
underdetermined dummy field $\mathbf \Omega^{(\kappa)}_{\rm dum}$.

The force on the electron beam can easily be calculated for $V \to
\infty$ by means of
 ~(\ref{derivative}) (with $\kappa=\alpha$) as a rapidly converging
sum involving Hankel 
functions. There is both a component in the backward direction as well
as a perpendicular component. The former is repulsive and invariant
under the replacement
$\alpha \to 1-\alpha$. The latter reverses sign under 
$\alpha \to 1-\alpha$ and under charge reversal, vanishes for
$\alpha=0,~1/2, ~1$, and for small $\Phi$
points in the same direction a Lorentz force would do for an electron
inside the cylinder.
For $kR \ll 1$ one obtains for the force per unit
cylinder length on an electron beam of
incoming momentum $\hbar k \mathbf e_x$ and density $\rho$
\begin{eqnarray}\label{force3}
\mathbf{F}^{(\Phi,V\to\infty)}= \rho \frac{\hbar^2k}{m_e}\left(
  \begin{array}{c} -2 \sin^2(\pi e\Phi/h) \\\sin(2\pi e\Phi/h)
  \end{array} \right)  + O(R^{\alpha'})
\end{eqnarray}  
where $\alpha'= {\rm min}(2\alpha,4-4\alpha)$.  The higher-order terms
in $R$ contain the reflecting force by the  cylinder and they vanish for
$R \to 0$ (magnetic string). For $kR \sim 1$, $F_2$ 
is several orders 
of magnitude smaller than $F_1$ since reflection by the cylinder
dominates. Subtracting this gives about the same order of magnitude,
but the remainder is overall much smaller than in  ~(\ref{force3}).

\section{Forces and the AB effect }
\label{sec_forces_and_AB}

These results yield a physical explanation of the AB effect as
follows. For the idealized case of an impenetrable cylinder,  the
quantum (but not the classical) dynamics  in the outside region is 
underdetermined
 since it contains a
largely arbitrary ``dummy'' 
field. Although tempting, there is no {\em a priori} reason to relate 
this field to an (in principle unknown) magnetic field
inside the cylinder. This leads to physically inequivalent Hamiltonians
$\hat{H}_{\rm dum}^{(\kappa)}$, $0 \leq\kappa < 1$, and a $\kappa$
dependent force. To derive an
additional boundary condition which removes the indeterminacy we
consider high, but 
finite, barriers. Then, by tunneling, minute tails of the scattering
wavefunction can enter the cylinder. The shielding and
a magnetic field inside the cylinder exert a force on these tails and
on the electron. By the combined influence of magnetic field and
increasing shielding 
on the form of the wavefunction this force remains  finite and flux dependent
even when the shielding 
goes to infinity and it can be expressed by the radial wavefunction derivative
at the cylinder. Requiring that the force for the directly tackled
idealized case (with 
infinite shielding from the outset) be equal to this limiting force
fixes the former's as yet underdetermined  dynamics
(\emph{i.e.} $\kappa$). Alternatively, the limit slope 
of the scattering solution at the boundary
can be considered as an additional boundary condition for the ideal
case which also removes the indeterminacy. The dummy
field determined in this way is, up to a factor $e$, just
the vector potential customarily used right 
away in the discussion of the AB effect.

\section{Summary}
It has been  shown that the AB effect for cylinders arises quite 
naturally when one considers tunneling and the force exerted on the small
wavefunction tails of the electron inside the cylinder. Although the
Lorentz force vanishes when the tails go to zero in 
the limit of infinite shielding,  the flux
dependence of the remaining force persists 
and precisely yields the AB effect. It has also been shown that the
limit slope of the scattering solution at the
boundary can be considered as an additional boundary condition for the
ideal case which also removes the indeterminacy of the quantization
procedure.
The same results are
expected to carry over to the torus and other domains.


\appendix\section{General case: Forces in a magnetic and scalar field}
\subsection{Magnetic and large, but finite, scalar potential}

We first consider a general time-independent magnetic field $\mathbf{B(x)}$
in a  region $G$,
with a vector potential $\mathbf{A(x)}$ vanishing at infinity, and a
scalar potential 
$V(\mathbf{x})$ which is nonzero in the same region $G$ and
vanishes outside.  Later we will remark on the more general case that
the scalar potential may also vanish on parts of the interior of $G$.

The force, ${\bf F}_t^{(V)}$, on an electron of charge $-e$ at time
$t$ is then the sum of the scalar  and Lorentz force,
\begin{equation}\label{forcea}
{\bf F}_t^{(V)} = \langle \psi_t^{(V)} , \left(- {\bf \nabla} V - e {\bf v}
  \times  {\bf B} \right) \psi_t^{(V)} \rangle 
\end{equation}
where $\psi_t^{(V)}$ denotes the wavefunction under the time
development with the Hamiltonian $H_V$, of an electron coming in from
infinity. The Hamiltonian is
\begin{equation}\label{ham}
H_V= ({\bf P}+ e {\bf A})^2/2m~+~V~.
\end{equation}
We will investigate the behavior of the force in
 ~(\ref{forcea}) for large scalar potential $V$ and will show that
it can be expressed as an integral over the surface, $\partial
G$, of the region $G$,
\begin{eqnarray}
{\bf F}_t^{(V)}&=& \frac{\hbar^2}{2m} \int_{\partial G} {\rm d} {\bf o}
\left| \frac{\partial \psi_t^{(V)}}{\partial n}
\right|^2 \label{a22}\\ 
&& ~~~~~~~~~~~~~~~~~~~~~~~+ ~{\rm terms}~ \to 0 ~~{\rm for}~~ V \to
\infty~. \nonumber 
\end{eqnarray}
Note that the dependence of this expression on the magnetic field
comes through the time development via the Hamiltonian. In two
dimensions and  if the region $G$  is a circle one arrives at
 ~(\ref{force1}) if one lets $\psi_t^{(V)}$ go to a stationary
solution of $H_V$.

To prove  ~(\ref{a22}) we first consider the scalar part in
 ~(\ref{forcea})  and show that 
\begin{eqnarray}
\langle \psi_t^{(V)},~ - \nabla V \psi_t^{(V)}\rangle &=& -
\frac{1}{2m} \int_G d^d x~ 
{\bf \nabla} \overline{\psi_t^{(V)}}  
{\bf P}^2 \psi_t^{(V)} + c. c. \nonumber \\
&&+ ~~{\rm terms}~ \to 0 ~~{\rm for}~ V \to \infty~. \label{a13}
\end{eqnarray}
To show this we use partial integration to write 
\begin{eqnarray}
\langle \psi_t^{(V)},~ - \nabla V \psi_t^{(V)}\rangle
&=&  \int d^d x~ V ({\bf x})~ {\bf \nabla}\!
\left| \psi_t^{(V)} ({\bf x}) \right|^2 \nonumber \\
&=& \int_G d^d x~ \nabla \overline{\psi_t^{(V)}} V
\psi_t^{(V)} + c.c. 
\label{a3}
\end{eqnarray}
$V$ can now be expressed by the Hamiltonian in   (\ref{ham}). For
the latter we have 
\begin{eqnarray}
&&\langle \psi_t^{(V)},H_V \psi_t^{(V)} \rangle \nonumber\\
&=&  \frac{1}{2m} \langle ({\bf P}+ e {\bf A}) \psi_t^{(V)},~({\bf P}+
e{\bf A}) \psi_t^{(V)} \rangle + \langle \psi_t^{(V)}, ~V\psi_t^{(V)}
\rangle \nonumber \\
&=&\frac{1}{2m} \langle ({\bf P}+ e {\bf A}) \psi_t^{(V)},~({\bf P}+
e{\bf A}) \psi_t^{(V)} \rangle _{I\!\!R^d\setminus G}\label{a5}\\
&&+
\frac{1}{2m} \langle ({\bf P}+ e {\bf A}) \psi_t^{(V)},~({\bf P}+
e{\bf A}) \psi_t^{(V)} \rangle_G + \langle \psi_t^{(V)}, ~V\psi_t^{(V)}
\rangle_G  \nonumber
\end{eqnarray}
where the indices  denote integration over the respective regions. The
term $\langle \psi_t^{(V)},H_V \psi_t^{(V)} \rangle$ is independent
of $t$ and therefore equals the incoming kinetic energy.  The
first term after the last equality sign converges for $V \to \infty$
to the corresponding expression with Dirichlet boundary conditions on
$G$, by standard results \cite{Davies,mv1995}, {\it e.g.} if ${\bf A}$ is
bounded and if 
the initial incoming wavefunction is in the domain of the Hamiltonian.
This limit also equals the incoming kinetic energy. Since all terms are
non-negative, the remaining terms in  ~(\ref{a5}) have to go to zero
when $V \to \infty$. Hence ${\rm for}~~V \to \infty$
\begin{eqnarray}
\langle\psi_t^{(V)}, H_V \psi_t^{(V)}\rangle_G && \to 0\label{a6}\\
\langle \psi_t^{(V)}, ~ V \psi_t^{(V)} \rangle\to 0,~~~~\langle
\psi_t^{(V)}, \psi_t^{(V)} \rangle_G && \to 
  0   \label{a7}\\
\langle ({\bf P} + e {\bf A}) \psi_t^{(V)},~ ({\bf P}+e {\bf A})
\psi_t^{(V)} \rangle_G && \to 0  \label{a8}
\end{eqnarray}
With this one obtains for bounded ${\bf A}$ and with the inequality
\begin{eqnarray}
\parallel ({\bf P}+e{\bf A}) \psi_t^{(V)} \parallel^2_G \geqslant
\left|\parallel {\bf P} \psi_t^{(V)} \parallel_G - \parallel \right. e
{\bf A} \left. \psi_t^{(V)} \parallel_G \right|^2 \label{a9} \nonumber
\end{eqnarray}
that also 
\begin{equation}
\parallel {\bf P} \psi_t^{(V)} \parallel_G \,\,\to 0~ ~~
{\rm for} ~~ V \to \infty~. \label{a10}
\end{equation}
This and Schwartz`s inequality then imply that
\begin{equation}
\langle {\bf \nabla} \psi_t^{(V)},~H_V \psi_t^{(V)} \rangle_G
  \to 0~~{\rm for}~~ V \to \infty \label{a11}
\end{equation}
if the initial wavefunction is in the domain of $H_V$. Inserting
$$
V = H_V - ({\bf P}^2 + e {\bf A})^2/2m
$$
into  ~(\ref{a3}) one obtains from  ~(\ref{a11})
\begin{eqnarray}
\langle \psi_t^{(V)}, - \nabla V \psi_t^{(V)}\rangle&=& - \frac{1}{2m}
\int_G d^d x~ {\bf \nabla} \overline{\psi_t^{(V)}} 
({\bf P} + e{\bf A})^2 \psi_t^{(V)} \nonumber \\
& & + c. c. \nonumber \\
&&+ ~~{\rm terms}~ \to 0 ~~{\rm for}~ V \to \infty~.
 \label{a12}
\end{eqnarray}
Using  ~(\ref{a10}) and Schwartz`s inequality this yields the claim
in  ~(\ref{a13}).

We will now investigate the first term on the right hand side of
 ~(\ref{a13}) and will show that it  can be
expressed as a surface integral, i.e. for the $i-$th component 
\begin{eqnarray}
- \frac{1}{2m} \int_G d^d x~ \!\!\!\!\!\!&& \partial_i
\overline{\psi_t^{(V)}} {\bf P}^2 \psi_t^{(V)} 
+ c.c. \nonumber\\ 
&=& \frac{\hbar^2}{2m} \left\{
\int_{\partial G} \partial_i \overline{\psi_t^{(V)}} {\bf \nabla}
\psi_t^{(V)} \cdot {\rm d}{\bf o} + c. c.\right. \nonumber\\
 &&~\left. -~~ {\bf e}_i \cdot \!\! \int_{\partial G}\!\!\!\!  
 {\rm d}{\bf o} ~~{\bf \nabla} \overline{\psi_t^{(V)}} \cdot {\bf
   \nabla}\psi_t^{(V)} 
\right\}~. \label{a15}
\end{eqnarray}
To prove this we use  partial integration, i.e. Gauss`s theorem in the
form
\begin{equation}
\int_G f \partial_j g = - \int_G (\partial_j f) g + \int_{\partial G}
{\bf e}_j \cdot {\rm d} {\bf o} \,  f\,g~. \label{a16}
\end{equation}
Then
\begin{eqnarray}
\!\!\!\!\!\!\int_G d^d x~ \partial_i \overline{\psi_t^{(V)}} \partial_j
\partial_j \psi_t^{(V)} &=& - \int_G d^d x~ \partial_j \partial_i
\overline{\psi_t^{(V)}} \partial_j \psi_t^{(V)}\nonumber\\
&& + \int_{\partial
  G}  {\bf e}_j \cdot {\rm d}{\bf o} \,\, \partial_i \overline{\psi_t^{(V)}}
\partial_j \psi_t^{(V)} \label{a17a}
\end{eqnarray}
and
\begin{eqnarray}
\!\!\!\!\!\!\int_G d^d x~ \partial_i \psi_t^{(V)} \partial_j \partial_j
\overline{\psi_t^{(V)}} &=& - \int_G d^d x~ {\partial_j \partial_i}
\psi_t^{(V)} \partial_j \overline{\psi_t^{(V)}}\nonumber\\
&& + \int_{\partial G} {\bf e}_j \cdot {\rm d}{\bf o}
 ~\partial_i \psi_t^{(V)} \partial_j
\overline{\psi_t^{(V)}}~. \nonumber \\
 \label{a17b}
\end{eqnarray}
Applying  ~(\ref{a16}) to  the first term on the r.h.s.\ of
 ~(\ref{a17b}) yields
\begin{eqnarray}
\!\!\!\!\!\!\int_G d^d x~ \partial_i \psi_t^{(V)} \partial_j \partial_j
\overline{\psi_t^{(V)}} &=& \int_G d^d x~ \partial_j \psi_t^{(V)} {\partial_i
\partial_j} \overline{\psi_t^{(V)}} \nonumber \\
& & - \int_{\partial G}   {\bf e}_i
\cdot {\rm d}{\bf o}~ \partial_j \psi_t^{(V)} \partial_j
\overline{\psi_t^{(V)}} \nonumber \\
& & + \int_{\partial G} {\bf e}_j \cdot {\rm d} {\bf o}
~\partial_i\psi_t^{(V)} \partial_j \overline{\psi_t^{(V)}}~. \nonumber
\\
\label{a17c}
\end{eqnarray}
Adding eqs.~(\ref{a17a}) and (\ref{a17c}), the first terms cancel and
one obtains  
\begin{eqnarray}
\!\!\!\!\!\!\int_G d^d x~ \partial_i \overline{\psi_t^{(V)}} \partial_j
\partial_j \psi_t^{(V)} + c.c. & & \nonumber \\
\!\!\!\!\!\! = \int_{\partial G} {\bf e}_j \cdot  {\rm d}{\bf o}~
\partial_i \overline{\psi_t^{(V)}} \partial_j \psi_t^{(V)} &+&
\int_{\partial  G} {\bf e}_j \cdot {\rm d}{\bf o}~ \partial_j
\overline{\psi_t^{(V)}} \partial_i \psi_t^{(V)}\nonumber\\
 &-& \int_{\partial G} {\bf e}_i \cdot  {\rm d}{\bf o}~
\partial_j \overline{\psi_t^{(V)}} ~\partial_j \psi_t^{(V)} \nonumber \\
\label{18}
\end{eqnarray}
which gives  ~(\ref{a15}).

For large $V$  ~(\ref{a15}) can be further simplified. Indeed, one
has $\psi_t^{(V)} \to 0$ on $\partial G$ for $V \to \infty$ and hence
\begin{eqnarray}
{\bf \nabla} \psi_t^{(V)} &\to& \frac{\partial \psi_t^{(V)}}{\partial n}
{\bf n} \qquad ({\bf n} = ~{\rm normal})\nonumber\\
\partial_i \psi_t^{(V)} &=& {\bf e}_i \cdot {\bf \nabla} \psi_t^{(V)} \to
{\bf e}_i \cdot {\bf n} \frac{\partial \psi_t^{(V)}}{\partial n} 
 \label{a19}
\end{eqnarray}
for $V \to \infty$. Insertion into  ~(\ref{a15}) gives 
\begin{eqnarray}
- \frac{1}{2m} \int_G d^d x~ \!\!\!\!\!\!&& \partial_i
\overline{\psi_t^{(V)}} {\bf P}^2 \psi_t^{(V)} 
+ c.c. \nonumber\\
&=& \frac{\hbar^2}{2m}   \int_{\partial G}  {\bf e}_i \cdot {\bf n}\,
{\rm do}
\frac{\partial \overline{\psi_t^{(V)}}}{\partial n} \frac{\partial
  \psi_t^{(V)}}{\partial n}  \label{a20}
\end{eqnarray}
plus terms going to 0 for $V \to \infty$.

For the Lorentz force  one has 
\begin{eqnarray}
\langle \psi_t^{(V)}, e{\bf B}  \times {\bf v} \psi_t^{(V)} \rangle
=
 \langle \psi_t^{(V)}, e{\bf B}
\times  m^{-1} ({\bf P}+ e {\bf A})
  \psi_t^{(V)}\rangle_G \nonumber
\end{eqnarray}
and one finds  from  ~(\ref{a10}) and Schwartz`s
inequality together with the boundedness of ${\bf B}$ that it
decreases to 0 when $V \to \infty\,$.
This, together with eqs.~(\ref{a20}) and (\ref{a13}) yields
 ~(\ref{a22}).

These results can be generalized to the situation when the potential
$V$ vanishes on a part 
of the interior of the region $G$, {\it e.g.} one could have a magnetic field in
a cylinder of radius $R$ while the shielding would be only in a
cylindrical ring $R_1 \leq r \leq R$. Then for particles coming in
from infinity one obtains the same results as above since the incoming
wavefunction will less and less penetrate through the ring. This can
be made mathematically precise by a similar reasoning as above.

\subsection{Forces in the case of infinite shielding}
If one starts right away with infinite shielding the particle motion
takes place completely outside the region $G$ and there is no regular
expression for the force acting at the boundary. We therefore use
${\rm d}/{\rm dt}~\langle m{\bf v} \rangle$ as definition for the
force. Moreover, the Hamiltonian is not uniquely determined, as
pointed out before, and we  arbitrarily pick for it a particular
$H_{\rm dum}$ with a dummy field $\mathbf \Omega_{\rm dum}$,
analogously to  ~(\ref{dummy}). Then
\begin{equation}\label{momentum}
m{\bf v} = {\bf P} + \mathbf \Omega_{\rm dum}.
\end{equation}
Now let $\psi_t$ be a normalizable
wavefunction coming in from infinity (and being in the domain of
$H_{\rm dum}$). We will show that then
\begin{equation}
\frac{\rm d}{\rm dt} \langle \psi_t, m {\bf v} \psi_t \rangle = 
\frac{\hbar^2}{2m} 
\int_{\partial G} {\rm d}{\bf o} 
\left| \frac{\partial \psi_t}{\partial n} \right|^2~. \label{b3}
\end{equation}
Note again that the dependence of this expression on the dummy
field $\mathbf \Omega_{\rm dum}$ comes through the time development
via the Hamiltonian. In two dimensions and  if the region $G$
 is a circle one
arrives at  ~(\ref{force2}) if one lets $\psi_t$ go to a stationary
solution of $H_{\rm dum}$.

To show  ~(\ref{b3}) we write the left hand side as 
\begin{equation}
\frac{\rm d}{\rm dt} \langle \psi_t, m {\bf v} \psi_t \rangle =
\frac{im}{\hbar} \big\{\langle H_{\rm dum} \psi_t, {\bf v} \psi_t
  \rangle - \langle \psi_t, {\bf v} H_{\rm dum} \psi_t \rangle
\big\} \label{b1}
\end{equation}
At this point it is {\it crucial} that in the first term  $H_{\rm dum}$
cannot be moved over to the other side by hermiticity 
 since then one would be led to $[H_{\rm dum}, {\bf v} ]={\bf
   \ddot{x}}= 0$  and thus there would be no force. The underlying
 mathematical reason why this is not allowed is that ${\bf 
  v} \psi_t$ need not lie in the domain of $H_{\rm dum}$.
However, one can move $\mathbf \Omega_{\rm dum} \cdot {\bf P}$ over by partial
integration since the boundary terms vanish. Thus
\begin{eqnarray}
\frac{\rm d}{\rm dt} \langle \psi_t, m {\bf v} \psi_t \rangle  & = &
\frac{im}{2m\hbar} \int_{I\!\!R^d\setminus G} d^d x~ \left\{  
\overline{{\bf P}^2 \psi_t} {\bf v}  \psi_t \right. \nonumber \\
& & \phantom{\overline{{\bf P}^2 \psi_t}} + \overline{ \psi_t}\,(\,2\,
 \mathbf \Omega_{\rm   dum}  \cdot
{\bf P} + \mathbf \Omega_{\rm   dum}^2)\, {\bf v}\, \psi_t \nonumber
\\
& & \left. \phantom{\overline{{\bf P}^2 \psi_t}} 
- \overline{\psi_t} {\bf v}
H_{\rm   dum} \psi_t \right\}~. \label{b2}
\end{eqnarray}
Inserting 
$0 = \overline{\psi_t} \hbar^2 {\bf
\nabla}^2 ({\bf v} \psi_t) - \overline{\psi_t} \hbar^2 {\bf \nabla}^2
({\bf v} \psi_t)$ 
gives
\begin{eqnarray*}
\frac{\rm d}{\rm dt} \langle \psi_t, m {\bf v} \psi_t \rangle 
&=& \frac{im}{2m \hbar} \int_{I\!\!R^d\setminus G} d^d x~ \left\{ 
- \hbar^2 ({\bf \nabla}^2 \overline{\psi_t}) {\bf v} \psi_t
\right. \nonumber \\
& & \phantom{\frac{im}{2m \hbar}} + \left.
\overline{\psi_t} \hbar^2 {\bf \nabla}^2 ({\bf v} \psi_t) +
\overline{\psi_t} [H_{\rm dum}, {\bf v}] \psi_t   \right\}~.
\end{eqnarray*}
Note that the last commutator vanishes. 
Inserting  ~(\ref{momentum})
 the $\mathbf \Omega_{\rm dum}$ terms
cancel, by partial integration as in  ~(\ref{a16}), since the
boundary terms vanish. Thus 
one obtains for the $i-$th  component of the force in  ~(\ref{b1})
\begin{eqnarray*}
\frac{\rm d}{\rm dt} \langle \psi_t, m  v_i \psi_t \rangle &=& \frac{-
  \hbar^2}{2m} \int_{I\!\!R^d\setminus G} d^d x~ 
\left\{ {\bf \nabla}^2 \overline{\psi_t} \partial_i \psi_t -
  \overline{\psi_t} {\bf \nabla}^2 \partial_i \psi_t\right\} \\
&&\!\!\!\!\!\!\!\!\!\!\!\!\!\!\!\!\!\!\!\!\!\!\!\!\!\!\!\!\!\!\!\!\!\!\!\!\!\!\!\!\!
 = \frac{- \hbar^2}{2m} \bigg\{
- \int_{I\!\!R^d\setminus G} d^d x~ {\bf \nabla} \overline{\psi_t} \cdot {\bf
  \nabla} \partial_i \psi_t - \int_{\partial {G}} {\rm d}{\bf o} \cdot {\bf
  \nabla} \overline{\psi_t} \partial_i \psi_t \\
& & \!\!\!\!\!\!\!\!\!\!\!\!\!\!\!\!\!\!\!\!\!\!\!\!\!\!\!\!\!\!\!\!\!\!
 + \int_{I\!\!R^d\setminus G} d^d x~ {\bf
  \nabla} \overline{\psi_t} \cdot {\bf \nabla} \partial_i \psi_t 
+ \int_{\partial {G} } {\rm d}{\bf o} \cdot \overline{\psi_t} {\bf
  \nabla} \partial_i \psi_t \bigg\}
 \end{eqnarray*}
where the last equality results from partial integration. 
The first and third term cancel, while the last one is zero since
$\psi$ vanishes on the boundary.
This yields
\begin{equation}
\frac{\rm d}{\rm dt} \langle \psi_t, m {\bf v} \psi_t \rangle =
\frac{ \hbar^2}{2m} \int_{\partial {G}} 
{\rm d}{\bf o} \cdot {\bf \nabla} \overline{\psi_t}{\bf
    \nabla}\psi_t~.
\end{equation}
Since $\nabla \psi_t$ is perpendicular to the boundary $\partial {G}$ this gives
 (\ref{b3}).


\end{document}